\documentclass[useAMS,usenatbib,referee]{biom}
\usepackage{amsmath}
\usepackage{graphicx,psfrag,epsf}
\usepackage{url} % not crucial - just used below for the URL %\usepackage{accents}
\usepackage{subcaption}
\newcommand{\bX}{\mathbf{X}}
\usepackage{comment}
\begin{document}
\title{Hypothesis Testing in Nonlinear Function on Scalar Regression with Application to Child Growth Study}
\author{Mityl Biswas \\ Department of Statistics, North Carolina State University, North Carolina, U.S.A 
\and 
Arnab Maity \\ Department of Statistics, North Carolina State University, North Carolina, U.S.A
\and
Jung-Ying Tzeng \\ Department of Statistics, North Carolina State University, North Carolina, U.S.A 
\and
Cathrine Hoyo \\ Department of Biological Sciences, North Carolina State University, North Carolina, U.S.A}
\pagerange{\pageref{firstpage}--\pageref{lastpage}} 
\label{firstpage}
\begin{abstract}
We propose a kernel machine based hypothesis testing procedure in nonlinear function-on-scalar regression model. The existing literature only deals with linear function-on-scalar regression models and there are no hypothesis tests performed in them. We use a kernel machine approach to model the unknown function and transform the hypothesis of no effect to an appropriate variance component test. Our research is motivated by the Newborn Epigenetic Study (NEST) which studies how maternal environmental exposures and nutrition affect child health. The question of interest is whether exposure to metals are associated with child growth. We take the child growth trajectory at a monthly resolution, which is a function of time, as the functional response, and model the measurements regarding metal content which are scalar covariates jointly using a nonlinear function. We demonstrate our proposed methodology using a simulation study and by applying it to analyze the NEST data.
\end{abstract}
\maketitle
\section{Introduction}\label{sec:intro}
In this article, we consider the problem of hypothesis testing in a functional regression model, where the responses are functions, possibly observed over an irregular set of points (e.g., time) and covariates are scalars. A popular model in this context is the linear function-on-scalar regression model, which models the covariates additively, and quantifies the effect of each of the covariates using a time-varying coefficient function. Our goal is to generalize such models to account for nonlinear effects of the covariates by jointly modeling them using an unknown, possibly nonlinear function. In this framework, we will develop a testing procedure to test for the association between the covariates and the functional response.

Our motivating data comes from the Newborn Epigenetic Study (NEST), a birth cohort study in Durham, North Carolina. Our goal is to determine the association between a child's growth trajectory for the first five years of its life and the level of its prenatal exposure to various metals. At birth, cord blood was collected and processed from babies to obtain the levels of prenatal exposure to multiple metals. The weight of the children was collected for several years from their birth. Our outcome of interest is the children's weights at a monthly level of resolution for the first five years, providing us between 2 and 30, randomly and irregularly spaced observations for each individual. Our goal is to investigate the association between the children's growth trajectories and their metal exposure levels. 

We face three main challenges. First, we wish to test the effect of the levels of multiple metals on growth trajectories. While testing for one metal at a time is straightforward, such an approach disregards any possible interactions and co-occurrences among the metals. Instead, we model the metals jointly. Second, we want to allow for the possibility that the effect on the growth trajectory may be nonlinear. Parametric models, such as linear or quadratic regression models, are not flexible enough; instead, we will use a semiparametric approach to model the joint effect of the metals. Finally, the functional data is relatively sparse and irregularly observed, and possibly with measurement errors.  These challenges make it difficult to analyze the data using existing functional methods. Our goal is to develop a new approach that can determine if there is a potentially nonlinear effect of a multidimensional covariate on irregularly-spaced functional data. 

Developing powerful statistical methods for studying the effect of prenatal exposure to metals is of significant interest in child development studies. Exposure to heavy metals, including cadmium, mercury, lead, arsenic, and zinc in the fetal stage has been found to cause low birth weights (\cite{Hu}, \cite{Sabra}). These metals are of great interest to the maternal reproductive health and fetal well being due to their ability to cross the placenta causing fetal toxicity (\cite{Hanna}, \cite{Clark}, \cite{Kampa}). While there is a plethora of literature investigating exposure to metals and child health, most of the articles discussed above focus only on one specific time point (e.g., weight at birth) or one feature of the growth pattern (e.g., velocity or acceleration). Most of the tests are performed for each metal one-at-a-time. Such an approach can not capture time-varying effects of metals on the growth curve, as well as any possible interactions among the metals that are ignored. In contrast, we consider the entire growth trajectory as the response and a group of metals as covariates. We aim to develop a testing procedure to investigate the association between the metals and the growth trajectory, taking the possible interactions and general non-linear effects into account. 

From a statistical point-of-view, we consider a function-on-scalar regression model, where we model the scalar covariates jointly using a nonlinear function. In the literature of functional data analysis, while there has been abundant work done in function-on-function and scalar-on-function regression, literature regarding function-on-scalar regression is much more scarce, with the research concentrated on linear models. The available articles focus on estimation problems and do not perform any hypothesis testing. With most of the articles concerning themselves with univariate covariates, the number of papers regarding multivariate covariates is low. Several Bayesian methods for function-on-scalar regression exist (\cite{Gold2}). \cite{Morris2} developed wavelet-based functional mixed models assuming that residual curves consist of independent measurement errors. \cite{Morris3} extended this to allow correlated residual curves. \cite{Golds} deals with multilevel functional responses, \cite{Gold2} estimates the bivariate function-on-scalar regression with subject level random functional effects while accounting for potential correlation in residual curves. A penalized spline approach for functional mixed models was taken in \cite{Bala}, whereas \cite{Bala2} used a piecewise constant basis. For cross-sectional functional data observed sparsely at the subject level, \cite{Montagna} developed a Bayesian latent factor model. The computational burden of the Bayesian procedures can be prohibitive for data exploration and model building even for moderate data sets, which has contributed to the slow adoption of Bayesian methods in functional data analysis. As an example, a comparison of the Bayesian penalized spline method in \cite{Bala} to a method based on functional principal component analysis on simulated data found computation times of 5 hours versus 5 seconds (\cite{Staicu}).

\cite{Faraway} considers the linear function-on-scalar model but his method works with an assumption that extremely few points are capable of representing an entire function, which is unreasonable. \cite{Barber} extends the LASSO to functional data. \cite{RHM} uses the same linear model as \cite{Faraway} to obtain a penalized generalized least squares(P-GLS) alternative. They have suggested a test statistic for hypothesis testing similar to an F statistic from \cite{Ramsay} without demonstrating it. \cite{Reiss} considers the same model as \cite{Faraway} and uses penalized regression with cubic B-spline basis function to avoid overfitting. They also suggest a hypothesis testing method using a pointwise F-statistic from (\cite{Ramsay}). However, how  well their suggested test statistic performs has not been demonstrated. \cite{RH} is similar to \cite{Reiss} in formulating the model, but uses pointwise restricted likelihood ratio test with tensor product penalty for smoothing bases. A specific method proposed by \cite{Wu} utilized kernel smoothing on the functional coefficients and assumed independent and identically distributed random errors. All of these articles deal with the linear model and would not be useful in dealing with a non-linear curve. They have also not suggested any method of performing hypothesis testing. \cite{LHZ} extends the functional single index model to a functional varying-coefficient single index model. The article also proposes, but does not illustrate a test statistic similar to the one by \cite{Reiss} for the same hypothesis test.

So far, all the articles dealing with function on scalar regression that exist in literature propose models that are linear in their scalar covariates. This may cause the model to fail to capture a non-linear relationship between the scalar covariate $\mathbf{X}$, and the functional response $\mathbf{y}(t)$. Pre-natal exposure to toxic metals has been cited in literature to be inversely associated with birth weight (\cite{Kippler}, \cite{Menai}, \cite{Vejrup}, \cite{Xie}, \cite{Jin}, \cite{He}, \cite{Rahman}, \cite{Uriu}), so a linear model would most likely fail to detect such a relationship. Moreover, they have not attempted to perform a hypothesis testing. In our approach we handle non-linear function on scalar regression, making it more universal. We have also proposed a method for testing of hypothesis, while comparing it to the hypothesis testing method proposed but not demonstrated in \cite{Ramsay}. Our method is capable of handling multivariate covariates.

The rest of the article is organized as follows: In section \ref{sec:meth} we explain the kind of data that motivates our work, and the way we model the data to get an appropriate hypothesis testing problem. We obtain a test statistic and use it to develop a test for the null hypothesis. Next, in section \ref{sec:sim} we have performed a simulation study for the cases of both dense data and sparse data. In this section, we have also compared the performance of our method with the one suggested by \cite{Reiss}. Subsequently, in section \ref{sec:dat} the methodology thus developed has been applied to the real-life dataset that motivated our work. Section \ref{sec:disc} discusses the results.
\section{Methodology}
\label{sec:meth}
\subsection{Framework and Model Specifications}
Suppose for the $i^{th}$ individual we observe data $Y_i(t_{ij})$, $X_{ik}$ and $Z_{i\ell}$ at $m_i$ many time points, $t_{ij}$ belonging to some compact interval for $j = 1, \ldots, m_i, i = 1, \ldots, n, k = 1, \ldots, p, \ell = 1, \ldots, q$. Here, $Y_i(t_{ij})$ is a functional response of time $t_{ij}$, $X_{ik}$ is the covariate of interest, and $Z_{i\ell}$ comprises the nuisance covariates. It is important to note that we are not imposing any restrictions on any of the aforementioned $X_{ik}$ or $Z_{i\ell}$.

To begin, let us consider the simpler situation of densely sampled data, where we have observations for $m$ data points for each individual, i.e., $m_i = m$ for $i = 1, \ldots, n$. Hence, $t_{ij} = t_j$ for $j = 1, \ldots, m.$ This assumption is only for illustration purposes, and we will discuss the sparsely sampled case which is more realistic in a later subsection. We consider the model
\begin{equation}
\label{1}
Y_i(t_{j}) = \sum_{\ell=1}^qZ_{i\ell}\eta_\ell(t_{j}) + \beta(\bX_i, t_{j}) + \varepsilon_i(t_{j}), 
\end{equation}
where $Y_i(t_{j})$ is the observed response corresponding to individual $i$ at time $t_{j}$, $Z_{i\ell}$ is the $\ell^{th}$ observed covariate which is not of interest to us (nuisance) for individual $i$, $\bX_i = (X_{i1}, \ldots, X_{ip})^T$ is the vector of observed covariates of interest to us corresponding to $Y_i(t_{j})$, at observed time $t_{j}$ for individual $i$, $\beta(\cdot, \cdot)$ is an unknown function determining how the observed response depends on the covariates of interest,  $\eta_\ell(t_{j})$ are the unknown regression coefficients of $Z_{i\ell}$, and $\varepsilon_i(t_{j})$ are the unknown error terms for the $i^{th}$ individual at time $t_{j}$, for $t_{j}$ belonging to some bounded continuous interval for, $i = 1, \ldots, n, j = 1, \ldots, m_i, \ell = 1, \ldots, q$. The $\epsilon_i(\cdot)$ are assumed to be Gaussian processes with mean $\mathbf{0}$, that are independent of each other, for $i = 1, \ldots, n$. We also assume that $\epsilon_i(\cdot)$ are independent of $\beta(\cdot, \cdot)$, $\bX_i$ and $Z_{i\ell}$ for $i = 1, \ldots, n, \ell = 1, \ldots, q$. $\bX_i$ and $Z_{i\ell}$ are assumed to be independent of each other for $i = 1, \ldots, n, \ell = 1, \ldots, q$. In order to account for the intercept term, we shall specify $Z_{i1} = 1, \text{for } i = 1, \ldots, n$. So, we have $q-1$ nuisance covariates for each person.

Our primary goal is to test whether there is any association between the covariate $\mathbf{X}_i$ and the functional response $\mathbf{Y}_i(\cdot)$. Thus, we aim to test the null hypothesis, $H_0: \beta(\cdot, \cdot) = 0$ against the alternate hypothesis $H_a: \beta(\cdot, \cdot) \neq 0$. Under $H_0$, the functional response $\mathbf{Y}_i(\cdot)$ would be independent of the vector covariate $\bX_i$. $\beta(\cdot, \cdot)$ is a bivariate function with a scalar operand and a vector operand. So, our problem is essentially an infinite-dimensional hypothesis testing problem.
\subsection{Hypothesis testing}
We want to test the null hypothesis, $H_0: \beta(\cdot, \cdot) = 0$. In this article, we adopt a Gaussian Process based approach. We assume $\beta(\cdot, \cdot)$ to be a Gaussian Process with mean $\mathbf{0}$, and variance-covariance kernel $\tau K(\cdot,\cdot)$, where $\tau\geq0$. This is a valid assumption to make if the $\epsilon_i(\cdot)$ have mean $\mathbf{0}$ for $i = 1, \ldots, n$, once we center $\mathbf{Y}_i(\cdot)$ about $\mathbf{0}$.

A Gaussian process is a process whose finite dimensional distributions are Gaussian. If the random function F is distributed as a Gaussian Process with mean function $\mu$ and covariance function $\kappa$, we denote it as $F \sim GP(\mu, \kappa)$. In Gaussian Process regression, given a set of input variables $\mathbf{W} = (w_1, w_2, \ldots, w_c)$, the latent function variable
$\mathbf{f} = (f_1,f_2, \ldots, f_c)$ has a joint Gaussian distribution
$P(\mathbf{f|W}) = N(\mathbf{M}, \mathbf{K_{GP}}(\mathbf{W}, \mathbf{W}))$,
where $\mathbf{M}$ is the mean vector with $c$ elements, and $\mathbf{K_{GP}}(\mathbf{W}, \mathbf{W})$ is a $c \times c$ covariance matrix (\cite{2}). The covariance matrix $\mathbf{K_{GP}}(\mathbf{W}, \mathbf{W})$ is in fact a kernel function evaluated at the $c$ instances. 

In our model, the observations for a given individual at different time points would be correlated to each other, and the observations for different individuals would also be correlated to each other at different time points. However, the covariance would be more when the observations are for the same individual. Also, for different individuals, the covariance would be more at time points that are closer together. To incorporate this, for two given time points, $t_h$ and $t_k$, we propose $\mathbf{K}$ as a matrix $\mathbf{K}\{(\mathbf{X}_i,t_h), (\mathbf{X}_j, t_k)\} = L(\mathbf{X}_i, \mathbf{X}_j)e^{-(t_{h}-t_{k})^2}$, where $L(\cdot, \cdot)$ is a kernel function, for $i = 1, \ldots, n, j = 1, \ldots, n, h = 1, \ldots, m, k = 1, \ldots, m$. Let us denote $\mathbf{t} = (t_1, \ldots, t_m)^T$. Then, for observed $((\mathbf{X}_1, \mathbf{t}), \ldots, (\mathbf{X}_n, \mathbf{t}))$,
\begin{center}
$(\beta(\mathbf{X}_1, \mathbf{t}), \ldots, \beta(\mathbf{X}_n, \mathbf{t})) \sim N(0, \tau \mathbf{K})$,
\end{center}
where $\mathbf{K}$ is the $mn \times mn$ kernel matrix with $n$ rows and $n$ columns of $m \times m$ blocks, with the $(i,j)^{th}$ block as $\mathbf{K}\{(\mathbf{X}_i,\mathbf{t}), (\mathbf{X}_j, \mathbf{t})\}$ which is a matrix with its $(h, k)^{th}$ element as $\mathbf{K}\{(\mathbf{X}_i,t_h), (\mathbf{X}_j, t_k)\}$. Such a $\mathbf{K}$ is clearly positive definite. So, $\beta(\cdot, \cdot) = 0$ is true if and only if $\tau = 0$. This provides us with the equivalent null hypothesis, $H_0': \tau = 0$, and the corresponding alternate hypothesis, $H_a': \tau \neq 0$. Then, the original problem of testing an infinite-dimensional function is converted to a much simpler one parameter test.

In the ideal scenario, $\eta_\ell(\cdot)$ would be known to us. However, in practice it is unknown. We model it using basis functions as $\eta_\ell(t_j) = \sum_{u = 1}^U B_{\ell u}(t_j)\theta_{\ell u}$, where $B_{\ell u}$ is a basis function, and $\theta_{\ell u}$ is the corresponding coefficient, for $j = 1, \ldots, m, \ell = 1, \ldots, q, u = 1, \ldots, U$. We use penalized likelihood maximization with thin plate regression splines as the basis functions. The smoothing parameters are estimated by generalized cross-validation criterion $n \frac{D}{(n - DoF)^2}$, where $D$ is the deviance, and $DoF$ is the effective degrees of freedom of the model. This lets us obtain estimates $\widehat{\eta}_\ell(\cdot)$ for $\ell = 1, \ldots, q$. The effect of the nuisance covariates is removed to get $Y^*_i(t_{j}) = Y_i(t_{j}) - \sum_{\ell=1}^qZ_{i\ell}\widehat{\eta}_\ell(t_{j})$.  
We obtain a test statistic from the score equation of this model to test whether $\tau=0$. We have developed a variance component test to incorporate the case of functional responses. 

Let $\Sigma(\cdot, \cdot)$ be the covariance operator of process $\varepsilon(\cdot)$, i.e., $\Sigma(s, t) = Cov(\varepsilon(s), \varepsilon(t))$. Let $\mathbf{R}$ be a matrix of dimension $m \times m$, with the $(j,k)^{th}$ element as $\Sigma(t_j, t_k)$. Let $\mathbf{\Sigma_0}$ be the block diagonal matrix of dimension $nm \times nm$ formed by $n$ blocks of $\mathbf{R}$. Then, $\mathbf{\Sigma_0}$ is the variance-covariance matrix of $\mathbf{\varepsilon} = (\varepsilon_1(t_1), \ldots, \varepsilon_1(t_m), \ldots, \varepsilon_n(t_1), \ldots, \varepsilon_n(t_m))$. 
Now, we have $Y^* \sim N(\mathbf{0}, \tau\mathbf{K} + \mathbf{\Sigma_0})$
This gives us the following likelihood function:
\begin{equation}
\label{6}
L(\tau|\mathbf{Y^*}, \mathbf{t}, \mathbf{X}) = e^{-\mathbf{Y^*}^T[\tau \mathbf{K} + \mathbf{\Sigma_0}]^{-1}\mathbf{Y^*} /2}\{|\tau \mathbf{K} + \mathbf{\Sigma_0}|^{1 / 2}(2\pi)^{nm / 2}\}^{-1}
\end{equation}
And the corresponding log-likelihood equation is obtained as $\ell(\tau|\mathbf{Y^*}, \mathbf{t}, \mathbf{X}) = -\mathbf{Y^*}^T[\tau \mathbf{K} + \mathbf{\Sigma_0}]^{-1}\mathbf{Y^*} /2 -log\{|\tau \mathbf{K} + \mathbf{\Sigma_0}|^{1 / 2}(2\pi)^{nm / 2}\}.$ When $\tau = 0$, we have \begin{align*}\frac{\delta \ell}{\delta \tau} = -\frac{1}{2}trace([\mathbf{\Sigma_0}]^{-1}\mathbf{K}) + \frac{1}{2} \mathbf{Y^*}^T[\mathbf{\Sigma_0}]^{-1}\mathbf{K}[\mathbf{\Sigma_0}]^{-1}\mathbf{Y^*}.\end{align*}
We note that only the second term depends on $\mathbf{Y^*}$. Thus, the test statistic we obtain is $T = \mathbf{Y^*}\mathbf{\Sigma_0}^{-1}\mathbf{K}\mathbf{\Sigma_0}^{-1}\mathbf{Y^*}^T$. 
However, the true error terms are not observable. So, we have to work with an estimate of $\mathbf{\varepsilon(\cdot)}$. From equation \ref{1} it follows that under $H_0$, $\widehat{\epsilon_i}(t_j) = Y_i^*(t_j)$ is an unbiased estimator of $\epsilon_i(t_j)$. Then, we can obtain an approximation of $\mathbf{\Sigma_0}$ as $\mathbf{\widehat{\Sigma}_0}$, where $\mathbf{\widehat{\Sigma}_0}$ is the variance-covariance matrix of $(Y^*_1(t_1), \ldots, Y^*_1(t_{m}), \ldots, Y^*_n(t_1), \ldots, Y^*_n(t_{m}))$. We consider the Karhunen-Loeve expansion of $\widehat{\varepsilon}_i(t_j)$, 
$\widehat{\varepsilon}_i(t_j) = \sum_{v = 1}^\infty\phi_v(t_j)\psi_{iv} + w_{ij}$,
where $w_{ij}$ is the white noise term, $w_{ij} \sim N(0, \sigma^2)$. This can be approximated by $\sum_{v = 1}^\zeta\phi_v(t_j)\psi_{iv} + w_{ij}$ 
for some suitable $\zeta$. Similarly, we can get the K-L expansion of $\widehat{\Sigma}(s, t)$, $\widehat{\Sigma}(s, t) = \sum_{v = 1}^\infty\phi_v(s)\phi_v(t)\lambda_v+ \sigma^2I(s = t)$.
This can be approximated by $\sum_{v = 1}^\zeta \phi_v(s)\phi_v(t)\lambda_v + \widehat{\sigma}^2I(s = t)$. We use functional principal component analysis to get $\zeta, \phi_v(t_j), \lambda_v$, $\widehat{\sigma}$ for $j = 1, \ldots, m, v = 1, \ldots, \zeta$. We obtain
$\widehat{R}_i(t_j, t_k) = \sum_{v = 1}^\zeta\phi_v(t_j)\phi_v(t_{k})\lambda_v + \widehat{\sigma}^2I(t_{j} = t_{k}),$
and we use it to construct $\mathbf{\widehat{\Sigma}_0}$.
The test statistic we propose is $T = \mathbf{Y^*}\mathbf{\widehat{\Sigma}_0}^{-1}\mathbf{K}\mathbf{\widehat{\Sigma}_0}^{-1}\mathbf{Y^*}^T$. If $\eta_\ell(\cdot)$ were known to us for $\ell = 1, \ldots, q$, then this test statistic would have a weighted sum of chi-squared distribution. But, we are estimating $\eta_\ell(\cdot)$ for $\ell = 1, \ldots, q$, so the test statistic does not have a standard distribution under the null hypothesis. Thus, in order to perform a valid test, the null distribution needs to be estimated. We use a re-sampling approximation permutation test (\cite{perm}). We proceed as follows for the $b^{th}$ iteration for $b = 1, \ldots, B$:
\begin{enumerate}
    \item Break the ordered sets of $(\mathbf{X}_i, \mathbf{Y}_i^*)$ into $(\mathbf{X}_i)$, and $(\mathbf{Y}_i^*)$ for all $i$ in $1, \ldots, n$.
    \item Some $(\mathbf{X}_{i^*})$ is randomly assigned to each $(\mathbf{Y}_i^*)$, where $\{i^*\}_1^n$ is a permutation of $\{1, \ldots, n\}$.
    \item The test statistic, $T^{(b)}$ is obtained for the set $(\mathbf{X}_{i^*}, \mathbf{Y}_i^*)$.
\end{enumerate}
  Under $H_0$, $\mathbf{X}$ and $\mathbf{Y^*}$ are unrelated, so the test statistic thus obtained in step $3$ is a valid test statistic under the assumption of the null hypothesis.  
  The distribution of the $B$ statistics thus obtained provides an empirical null distribution of the test statistics. 
However, here we remove the relation between $\mathbf{X}$ and $\mathbf{Z}$ as well. So, a test based on this would be stricter than required. 

If $H_0:\tau=0$ is rejected at desired level of significance, $\alpha$, then we select the covariate $\mathbf{X}$ as a predictor for the variable $\mathbf{Y}$. The p-value is estimated as the proportion of times we obtain a test statistic as extreme as the one observed on simulating the test statistic a large number of times from the null model. We estimate it as $\frac{\sum_{b = 1}^B I(T^{(b)} > T) + 1}{B + 1}$. Note that the performance of our test statistic would depend on the choice of the kernel function $L(\cdot, \cdot)$. We have investigated which commonly used kernel might be preferred through simulations in section \ref{sec:sim}.
\subsection{Extension to Sparsely Sampled Data}
The discussion so far corresponded to an ideal case where the data obtained is dense. However, real data is generally sparse. So, we need to make certain adjustments. We consider the following model, which is an extension of equation $\ref{1}$:
\begin{equation}
\label{2}
Y_i(t_{ij}) = \sum_{\ell=1}^qZ_{i\ell}\eta_\ell(t_{ij}) + \beta(\bX_i, t_{ij}) + \varepsilon_i(t_{ij}), 
\end{equation}
where $Y_i(t_{ij})$ is the observed response corresponding to individual $i$ at time $t_{ij}$, $Z_{i\ell}$ is the $\ell^{th}$ observed covariate which is not of interest to us (nuisance) for individual $i$, $\bX_i = (X_{i1}, \ldots, X_{ip})^T$, is the vector of observed covariates of interest to us corresponding to $Y_i(t_{ij})$, at time $t_{ij}$ for individual $i$, $\beta(\cdot, \cdot)$ is a function determining how the observed response depends on the covariates of interest,  $\eta_\ell(t_{ij})$ are the regression coefficients of $Z_{i\ell}$, and $\varepsilon_i(t_{ij})$ are error terms for the $i^{th}$ individual at time $t_{ij}$, for $t_{ij}$ belonging to some bounded continuous interval, $i = 1, \ldots, n, j = 1, \ldots, m_i, \ell = 1, \ldots, q$. The $\epsilon_i(\cdot)$ are assumed to be Gaussian processes with mean $\mathbf{0}$, that are independent of each other, for $i = 1, \ldots, n$. We also assume that $\epsilon_i(\cdot)$ are independent of $\beta(\cdot, \cdot)$, $\bX_i$ and $Z_{i\ell}$ for $i = 1, \ldots, n, \ell = 1, \ldots, q$. $\bX_i$ and $Z_{i\ell}$ are assumed to be independent of each other for $\ell = 1, \ldots, q$. In order to account for the intercept term, we shall specify $Z_{i1} = 1$, for $i = 1, \ldots, n$. We use this framework to deal with the same hypothesis testing problem as before.

We extend the approach we made for densely sampled data to sparsely sampled data. $K$ is now defined as $K\{(\mathbf{X}_i, t_{ih}), ( \mathbf{X}_j, t_{jk})\} = L(\mathbf{X}_i, \mathbf{X}_j)e^{-(t_{ih}-t_{jk})^2}$, where $L(\cdot, \cdot)$ is a kernel function as in the densely sampled scenario, for $i = 1, \ldots, n, j = 1, \ldots, n, h = 1, \ldots, m_i, k = 1, \ldots, m_j$. Therefore, for observed $\{(\mathbf{X}_1, t_{11}), \ldots, (\mathbf{X}_n, t_{nm_n})\}$, $\{\beta(\mathbf{X}_1, t_{11}), \ldots, \beta(\mathbf{X}_n, t_{nn})\} \sim N(0, \tau \mathbf{K})$, where $\mathbf{K}$ is the $\sum_{i = 1}^nm_i\times\sum_{i = 1}^nm_i$ kernel matrix with $n$ rows and $n$ columns of $m_i \times m_j$ blocks, for $i = 1, \ldots, n, j = 1, \ldots, n$, with the $(i,j)^{th}$ block as $\mathbf{K}\{(\mathbf{X}_i, t_i), (\mathbf{X}_j, t_j)\}$ which is an $m_i \times m_j$ matrix with its $(h, k)^{th}$ element as $\mathbf{K}\{(\mathbf{X}_i, t_{ih}), (\mathbf{X}_j, t_{jk})\}$. $\eta_\ell(t_{ij})$ is estimated by generalized additive modeling of $Y_i(t_{ij})$ on $Z_{i\ell}$ to get $\widehat{\eta}_\ell(t_{ij})$ for $i = 1, \ldots, n, j = 1, \ldots, m_i, \ell = 1, \ldots, q$. We model $\eta_\ell(t_{ij}) = \sum_{u = 1}^U B_{\ell u}(t_{ij})\theta_{\ell u}$, where $B_{\ell u}$ is a basis function, and $\theta_{\ell u}$ is the corresponding coefficient, for $i = 1, \ldots, n, j = 1, \ldots, m_i, \ell = 1, \ldots, q, u = 1, \ldots, U$. The effect of the nuisance covariates is removed to get $Y^*_i(t_{ij}) = Y_i(t_{ij}) - \sum_{\ell=1}^qZ_{i\ell}\widehat{\eta}_\ell(t_{ij})$.
$\mathbf{R}_i$ is a matrix of dimension $m_i \times m_i$, with the $(j,k)^{th}$ element as $\Sigma(t_{ij}, t_{ik})$. Let $\mathbf{\Sigma_0}$ be the block diagonal matrix of dimension $\sum_{i = 1}^nm_i \times \sum_{i = 1}^nm_i$ formed by $\mathbf{R}_1, \ldots, \mathbf{R}_n$. The variance-covariance matrix of $\mathbf{\varepsilon} = \{\varepsilon_1(t_{11}), \ldots, \varepsilon_1(t_{1m_1}), \ldots, \varepsilon_n(t_{n1}), \ldots, \varepsilon_n(t_{nm_n})\}$ is $\mathbf{\Sigma_0}$. But the true error terms being unobservable, we estimate $\mathbf{\varepsilon}$ by $\widehat{\epsilon_i}(t_{ij}) = Y_i^*(t_{ij})$ which is an unbiased estimate under the null hypothesis. The variance-covariance matrix of $\{Y^*_1(t_{11}), \ldots, Y^*_1(t_{1m_1}), \ldots, Y^*_n(t_{n1}), \ldots, Y^*_n(t_{nm_n})\}$ is $\mathbf{\widehat{\Sigma}_0}$. We consider the Karhunen-Loeve expansion of $\varepsilon_i(t_{ij})$ and $\Sigma(s, t)$ and use approximate the latter by $\sum_{v = 1}^\zeta\phi_v(s)\phi_v(t)\lambda_v + \sigma^2I(s = t)$. We use functional principal component analysis to get $\zeta, \phi_v(t_{ij}), \lambda_v$ for $j = 1, \ldots, m_i, v = 1, \ldots, \zeta$. 
We obtain $\mathbf{\widehat{R}}_i(t_{ij}, t_{i\ell}) = \sum_{k = 1}^\zeta\phi_k(t_{ij})\phi_k(t_{i\ell})\lambda_k + \sigma^2I(t_{ij} = t_{i\ell})$, and use it to construct $\mathbf{\widehat{\Sigma}_0}$.
The likelihood function is similar to equation \ref{6}, with $nm$ replaced by $\Sigma_{i = 1}^n m_i$.
This leads to similar changes in the subsequent equations, eventually giving us the same test statistics.
The permutation based method is similar to the one for the case of densely sampled data. 
\section{Simulation Study}
\label{sec:sim}
\subsection{Design of Study}
We performed a simulation study to evaluate the finite-sample performance of our test and compared its performance with the F-like statistic suggested in \cite{Reiss}. For testing the performance of the method in terms of type I error, the data sets were simulated $s = 10,000$ times from the true model \ref{2}, under the null hypothesis $\beta(\cdot, \cdot) = 0$, with $n = 100$ subjects, $p = 5$ covariates of interest for each subject, and $q = 3$ nuisance covariates for each subject. Note that model \ref{1} is a special case of model \ref{2}. For $i = 1, \ldots, n$, and $j = 1, \ldots, m_i$, the following was done: The covariates of interest were generated as $X_{ij} = \mathbf{X_i} \sim N_q(\mathbf{0}, \mathbf{I_q})$, and the nuisance covariates were generated as $Z_{i1} = 1$, $Z_{i2} \sim N(0, 1)$, $Z_{i3} \sim Ber(0.4)$, with their corresponding coefficients in the model being $\eta_1(t) = t$, $\eta_2(t) = sin(2\pi t)$, $\eta_3(t) = cos(2\pi t)$. The error functions were generated as $\varepsilon(t_{ij}) = \sqrt{2}a_{i1}cos(2\pi t_{ij}) + \sqrt{2}a_{i2}sin(2\pi t_{ij}) + w_{ij}$, where $a_{i1} \sim N(0, 1)$, is independent of $a_{i2} \sim N(0, 2)$, and they are both independent of $w_{ij} \sim N(0, 1)$. The other specifications vary based on whether the simulations were for densely sampled or sparsely sampled data, and are as follows:
\begin{enumerate}
\item Densely sampled data
We considered $m_i=m=51$, $t_{ij} = \frac{j}{m}, i = 1, \ldots, n, j = 1, \ldots, m$.
The values of $\delta$ used were $\delta_1 = (0, 0.1, 0.2, \ldots, 1), \delta_2 = (0, 0.1, 0.2, \ldots, 1.2), \delta_3 = (0, 0.1, 0.2, \ldots \\, 1),$ and $\delta_4 = (0, 0.1, 0.2, \ldots, 1)$  for $\beta_1(\cdot, \cdot), \beta_2(\cdot, \cdot), \beta_3(\cdot, \cdot),$ and $\beta_4(\cdot, \cdot)$, respectively. 
\item Sparsely sampled data
We generated $m_i$ $\epsilon$ $\{7, 8, \ldots, 14\}$ with equal probability, and considered $t_{ij} = \frac{j}{m}$. Then, $t_{ik}$ was obtained through simple random sampling without replacement of size $m_i$ from $t_{ij}$ for $i = 1, \ldots, n, j = 1, \ldots, m, k = 1, \ldots, m_i$.
The values of $\delta$ used were $\delta_1 = (0, 0.3, 0.6, \ldots, 1.5), \delta_2 = (0, 0.5, 1, \ldots, 4), \delta_3 = (0, 0.3, 0.6, \ldots \\, 1.5)$, and $\delta_4 = (0, 0.5, 1, \ldots, 3)$ for $\beta_1(\cdot, \cdot), \beta_2(\cdot, \cdot), \beta_3(\cdot, \cdot),$ and $\beta_4(\cdot, \cdot)$, respectively.
\end{enumerate}
We compared the proposed test with the statistic similar to the one suggested in \cite{Reiss}, which is henceforth referred to as the F-statistic. For this purpose, we considered the model $Y_i(t_{ij}) = \sum_{\ell=1}^qZ_{i\ell}\eta_\ell(t_{ij}) + \mathbf{x_i}^T\gamma(t_{ij}) + \varepsilon_i(t_{ij})$ for $j = \{1, \ldots, m_i\}, i = \{1, \ldots, n\}$. The null model had $\gamma(t_{ij}) = 0$, for $j = \{1, \ldots, m_i\}, i = \{1, \ldots, n\}$, and under $H_a$, $\gamma(t_{ij})$ and $\eta_\ell(t_{ij})$ were estimated by the same procedure used to estimate $\eta_\ell(t_{ij})$ in our own methodology, for $j = \{1, \ldots, m_i\}, i = \{1, \ldots, n\}$. The test statistic used was $F^* = \frac{[RSS_0 - RSS_1]}{RSS_1}$ where, $RSS_0$ and $RSS_1$ are the residual sum of squares under $H_0$ and $H_a$, respectively.

In each scenario, we compare the performance of three different kernel functions, $L(\cdot,\cdot)$: $L_1(\cdot,\cdot)$ the linear kernel, $L_2(\cdot,\cdot)$ the quadratic kernel, and $L_3(\cdot,\cdot)$ the gaussian kernel. The $d^{th}$ polynomial kernel corresponds to the models with $d^{th}$-order polynomials including the cross-product terms (\cite{Maity}). The Gaussian kernel generates the function space spanned by the radial basis functions (\cite{Buhmann}) and incorporates several nonlinear functions.  
The null distribution was simulated by the permutation method using $B = 1,000$ permutations. This gave us $B$ simulations of the test statistics for each case, namely $T_{ij1}, \ldots, T_{ijB}$ for the $i^{th}$ scenario and the $j^{th}$ simulation for $i = 1, \ldots, 4, j = 1, \ldots, s$. For an observed test statistic $T^*_{ij}$ in the $i^{th}$ scenario and $j^{th}$ simulation, its estimated p-value is $p_{ij} = \frac{\sum_{k = 1}^B I(T_{ijk} > T_{ij}^*) + 1}{B + 1}$ for $i = 1, \ldots, 4, j = 1, \ldots, s$. The type I error at nominal level $\alpha$ was estimated as $\frac{\sum_{j = 1}^s I(p_{ij} > \alpha) + 1}{s + 1}$ for each of our four scenarios $i = 1, \ldots, 4$ for $\alpha_1 = 0.01$, $\alpha_2 = 0.05$, and $\alpha_3 = 0.1$. 

The power was computed at $\alpha_{pow} = 0.05$ level of type I error. We performed $s_{pow} = 400$ simulations under the same setup as the one used for simulating type-I error, with $B_{pow} = 400$ permutations, but with different values of $\beta(\cdot,\cdot)$. For the purpose of estimating the power, $\beta_1(\mathbf{X}_i, t) = \delta\overline{\mathbf{X_i}}t$, $\beta_2(\mathbf{X}_i, t) = \delta\overline{\mathbf{X_i}}^2t$, $\beta_3(\mathbf{X}_i, t) = \delta e^{-\overline{\mathbf{X_i}}t}$, and $\beta_4(\mathbf{X}_i, t) = \delta e^{-\overline{\mathbf{X_i}^2}t}$ were considered, where $\overline{\mathbf{X_i}} = \frac{1}{p}\sum_{k=1}^pX_{ik},$ and $\mathbf{\overline{X_i}}^2 = \frac{1}{p}(\sum_{k=1}^pX_{ik})^2$. The linear case was demonstrated by $\beta_1(\mathbf{X}_i, t)$, the quadratic case by $\beta_2(\mathbf{X}_i, t)$, and non-separable cases by $\beta_3(\mathbf{X}_i, t)$ and $\beta_4(\mathbf{X}_i, t)$. Due to the fact that $\beta_3(\mathbf{X}_i, t)$ may be approximated by a linear function from its Taylor series expansion in certain cases, $\beta_4(\mathbf{X}_i, t)$ has also been considered. The power was estimated as $P = \frac{\sum_{j = 1}^s I(p_{ij} < \alpha_{pow})}{s}$ for $i = 1, \ldots, 4$. Different $\delta$ values were considered to determine where sufficient power was being achieved. It would be expected that the power would be lower when the $\delta$ value is lower, and when the data is sparse.  
Power curves were drawn for the purpose of comparison. A higher power at a lower value of $\delta$ would make the corresponding kernel more useful.
\subsection{Results}
\label{res}
The results from the simulation study are presented in Table~\ref{1} for the type I error, and Figures 1 and 2 for the power analysis for the densely sampled case and the sparsely sampled case, respectively. Our method appears to have type I error around the nominal level. 
The relative performances of the kernels in the dense scenario are similar to the relative performances of the kernels in the sparse scenario. Linear kernel gives higher power for $\beta_1$, and quadratic kernel gives higher power for $\beta_2$. This is as would be expected. Linear kernel also gives higher power for $\beta_3$, this would be as the linear term is the most significant in the Taylor series expansion of the exponential term. For $\beta_4$, the Gaussian kernel gives the most power. We see that a linear kernel is incapable of detecting the presence of a non-linear component. It is observed that the Gaussian kernel works better than the linear kernel for non-linear $\beta$. While the quadratic kernel is better when $\beta$ is quadratic, when $\beta$ is a more complicated function, like the non-separable $\beta_4$, the Gaussian kernel performs best.\\
\begin{table}
\label{t:1}
\resizebox{\textwidth}{!}
{\begin{minipage}{\textwidth}
\begin{tabular}[htbp]{|c|c|c|c|c|}
        \hline
        \multicolumn{5}{|c|}{\textbf{Dense case}} \\
        \hline
        \textbf{Nominal value of $\alpha$}&{\textbf{Linear}}  & {\textbf{Quadratic}} & {\textbf{Gaussian}} & \textbf{F-statistic}\\
        \hline
        \textbf{0.01} & 0.011 & 0.011 & 0.009 & 0.012\\
        \hline
        \textbf{0.05} & 0.054 & 0.051 & 0.047 & 0.051 \\
        \hline
        \textbf{0.1} & 0.101 & 0.102 & 0.100 & 0.100 \\
        \hline
        \multicolumn{5}{|c|}{\textbf{Sparse case}} \\
        \hline
        \textbf{Nominal value of $\alpha$}&{\textbf{Linear}}  & {\textbf{Quadratic}} & {\textbf{Gaussian}} & \textbf{F-statistic}\\
        \hline
         \textbf{0.01} & 0.010 & 0.008 & 0.010 & 0.011\\
        \hline
        \textbf{0.05} & 0.050 & 0.048 & 0.049 & 0.051 \\
        \hline
        \textbf{0.1} & 0.096 & 0.098 & 0.099 & 0.102 \\
        \hline 
        \end{tabular}
        \bigskip
        \caption{Results for simulation study described in section \ref{sec:sim}. Displayed are the type I error rates (rounded to third decimal place) for both the dense and sparse cases for all three kernels considered and the F-statistic.}
\end{minipage}}
\end{table} 
\begin{figure}[ht]
\begin{tabular}{ll}
\includegraphics[scale=0.45]{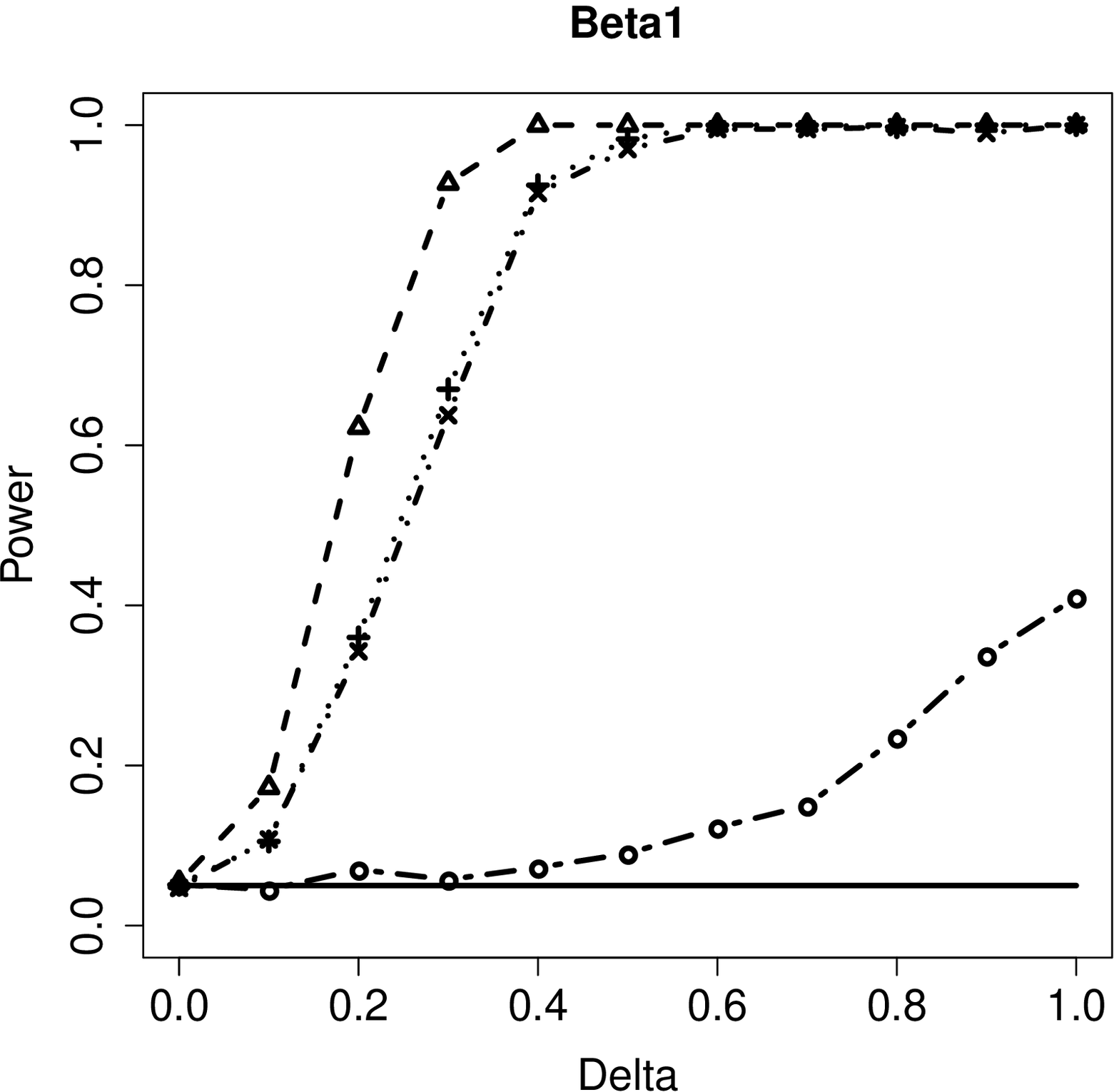} &
\includegraphics[scale=0.45]{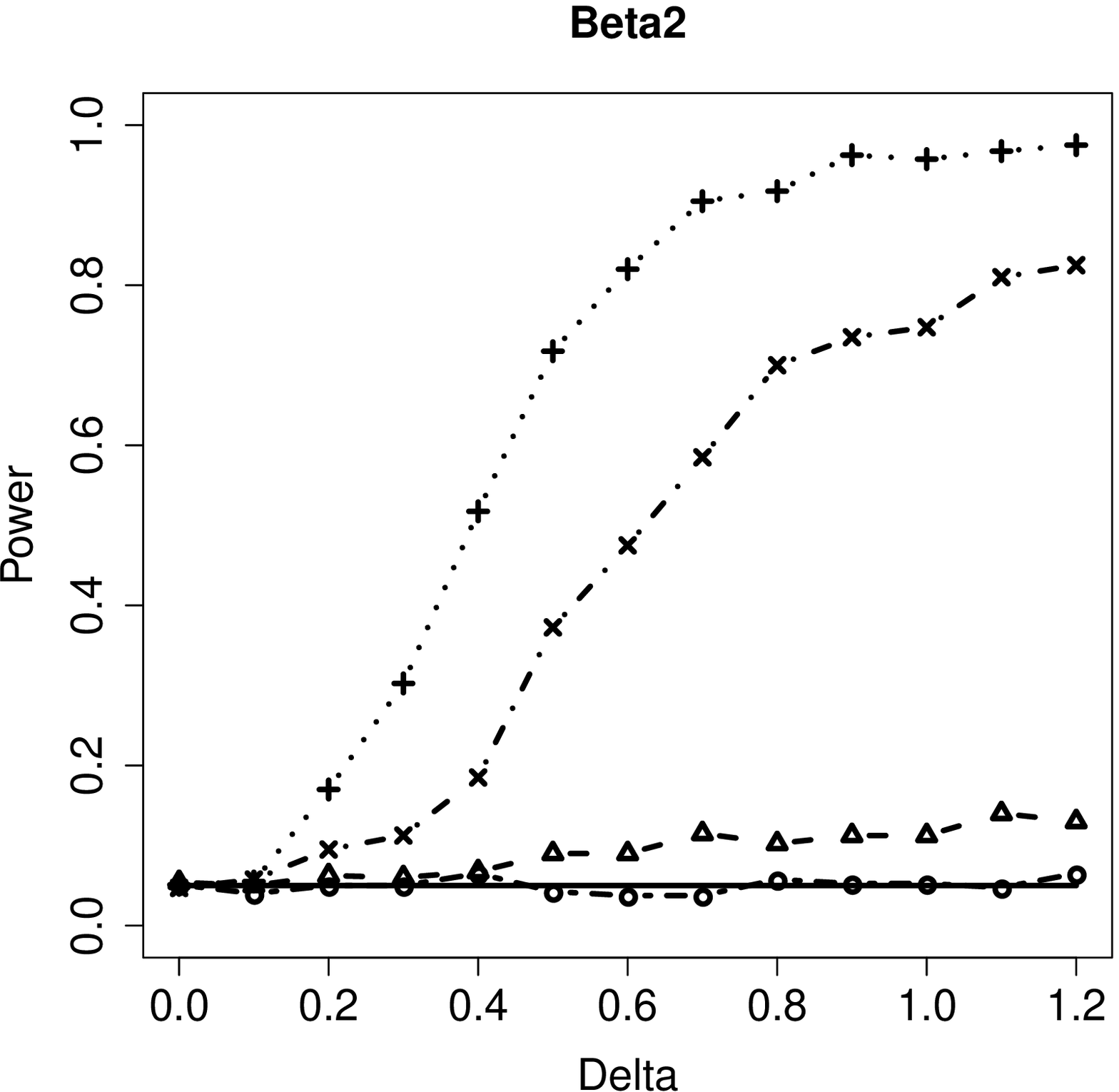}\\
\includegraphics[scale=0.45]{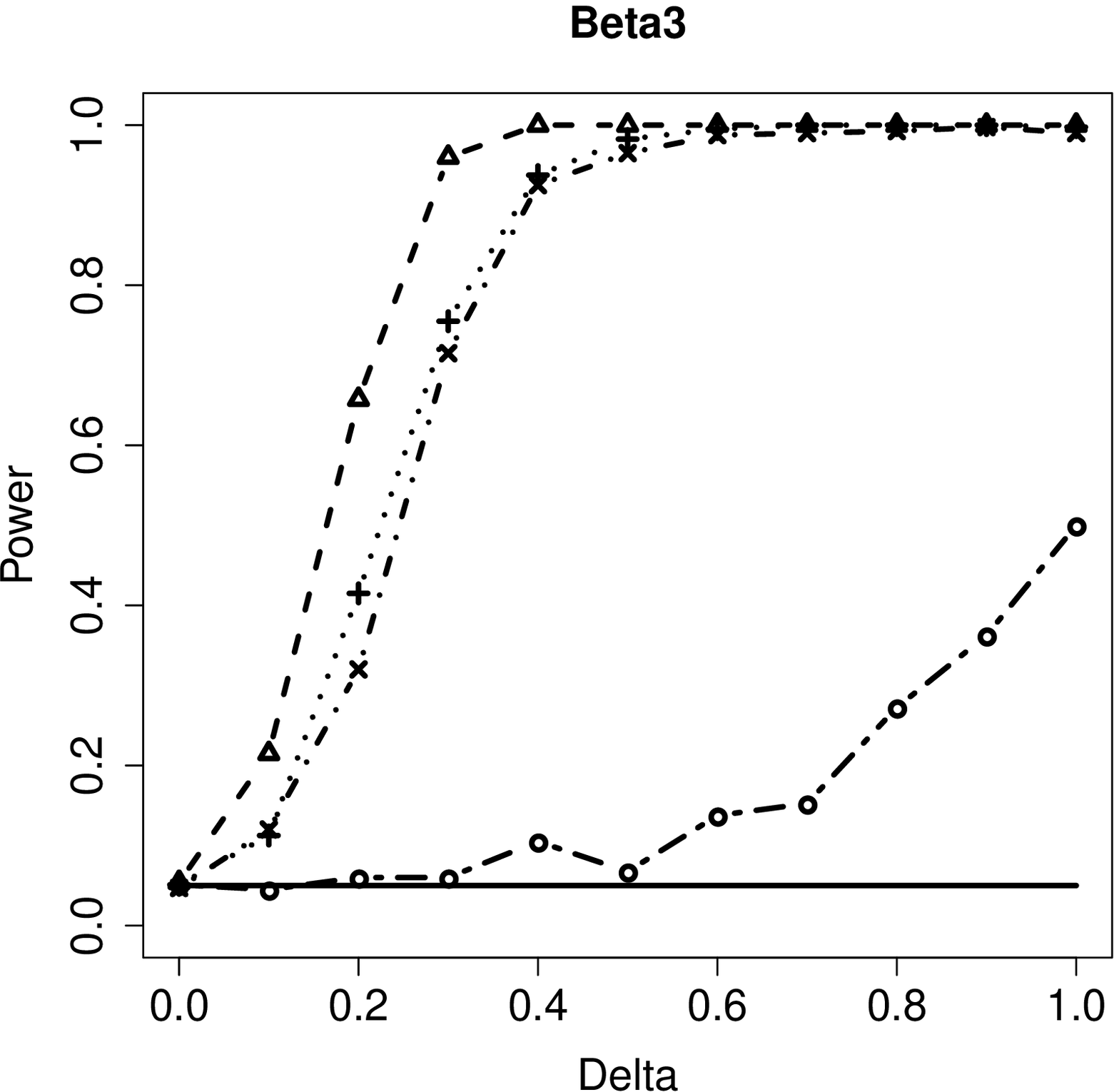} &
\includegraphics[scale=0.45]{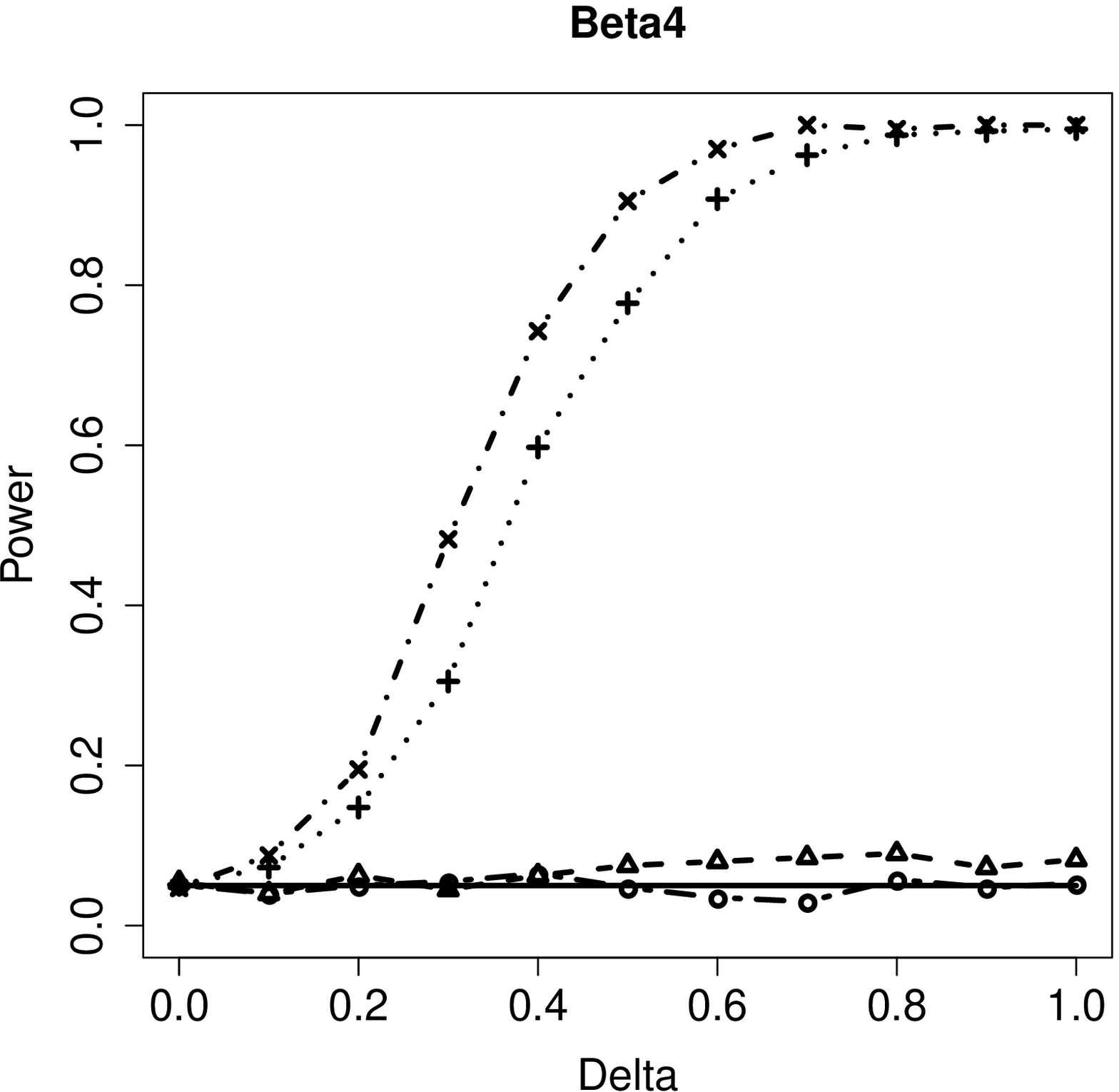}\\
&
\end{tabular}
\caption{Results illustrating the power curves from the simulation study described in section \ref{sec:sim} for $T^*$, and the F-statistic for densely sampled data for $\beta_1(\cdot, \cdot)$, $\beta_2(\cdot, \cdot)$, $\beta_3(\cdot, \cdot)$, and $\beta_4(\cdot, \cdot)$ (Linear kernel : dashed line with triangles, Quadratic kernel : dotted line with pluses, Gaussian kernel : dash and dot line with crosses, F-statistic : small and big dash with circles, solid horizontal line at nominal level of $\alpha, 0.05$).}
\end{figure}
%% Fig 2
\begin{figure}[ht]
\begin{tabular}{ll}
\includegraphics[scale=0.45]{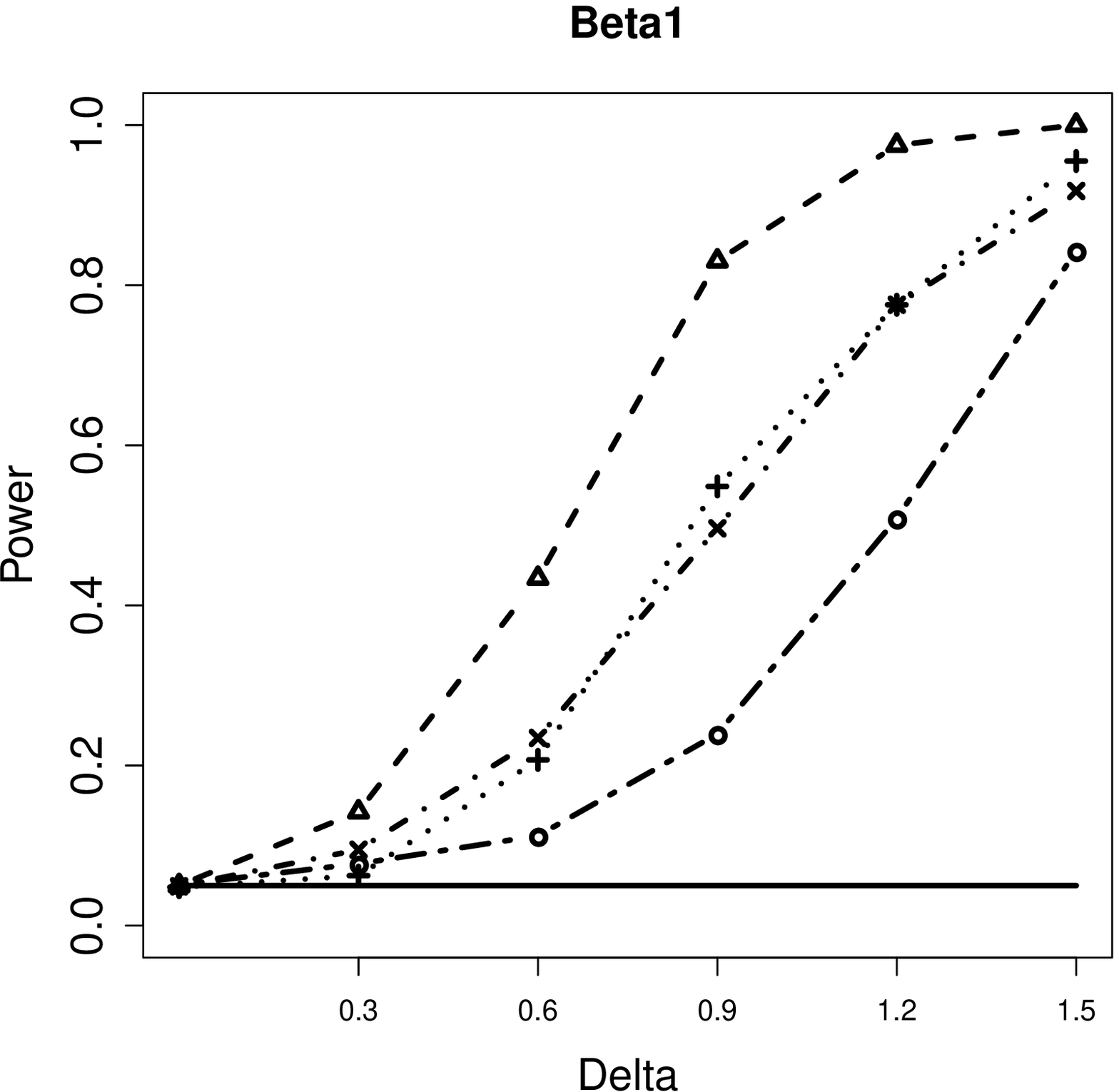} &
\includegraphics[scale=0.45]{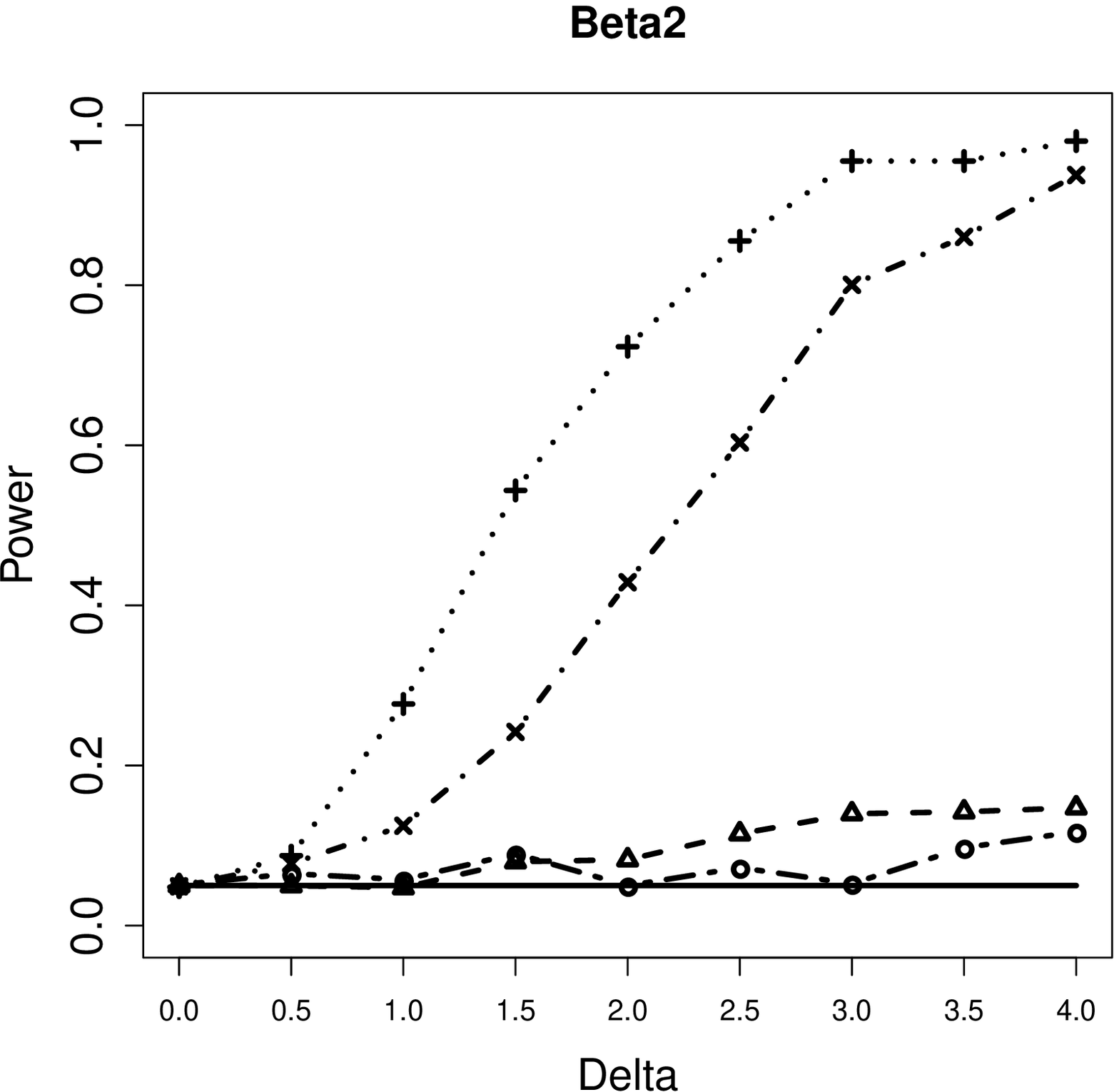}\\
\includegraphics[scale=0.45]{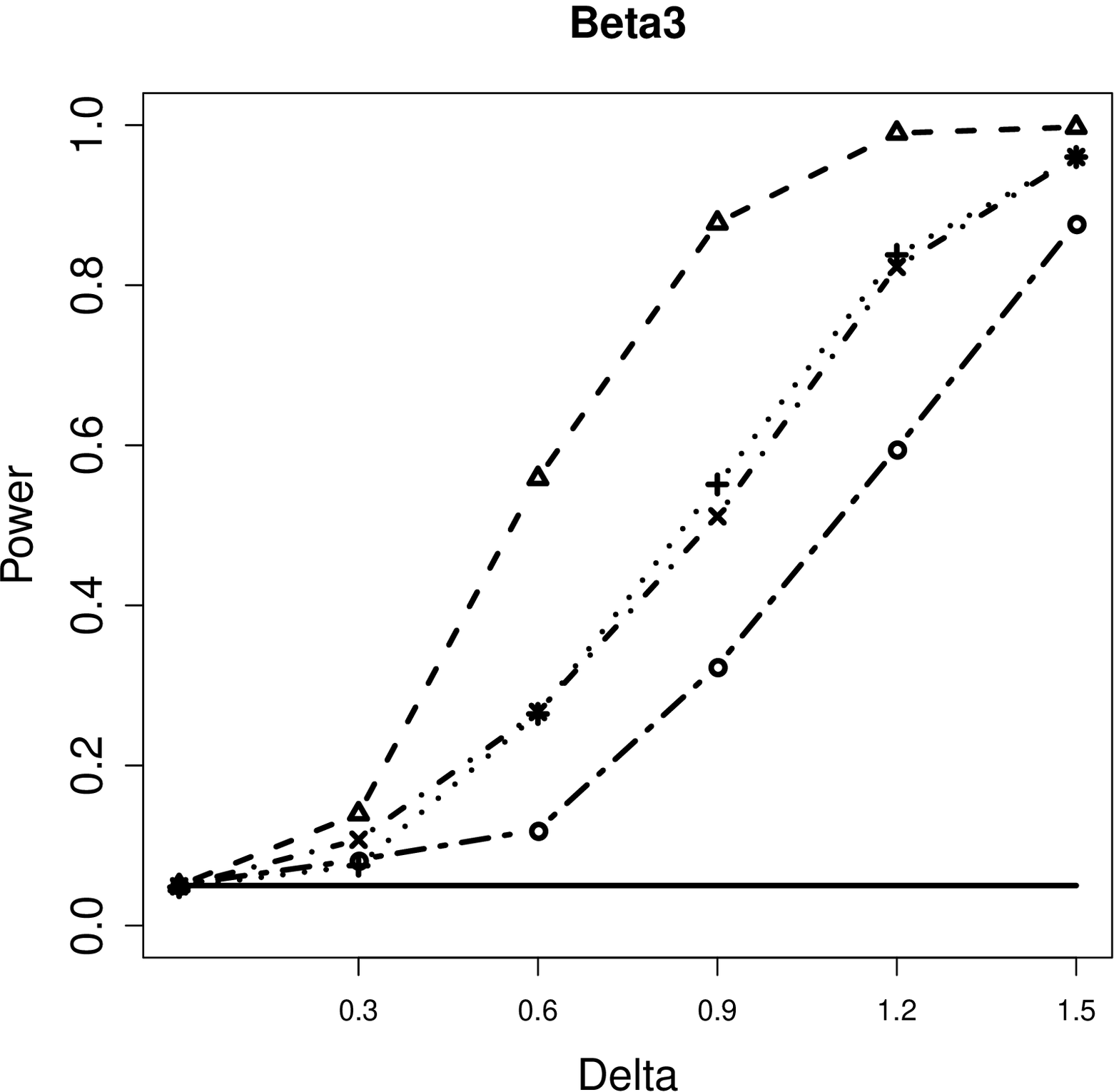} &
\includegraphics[scale=0.45]{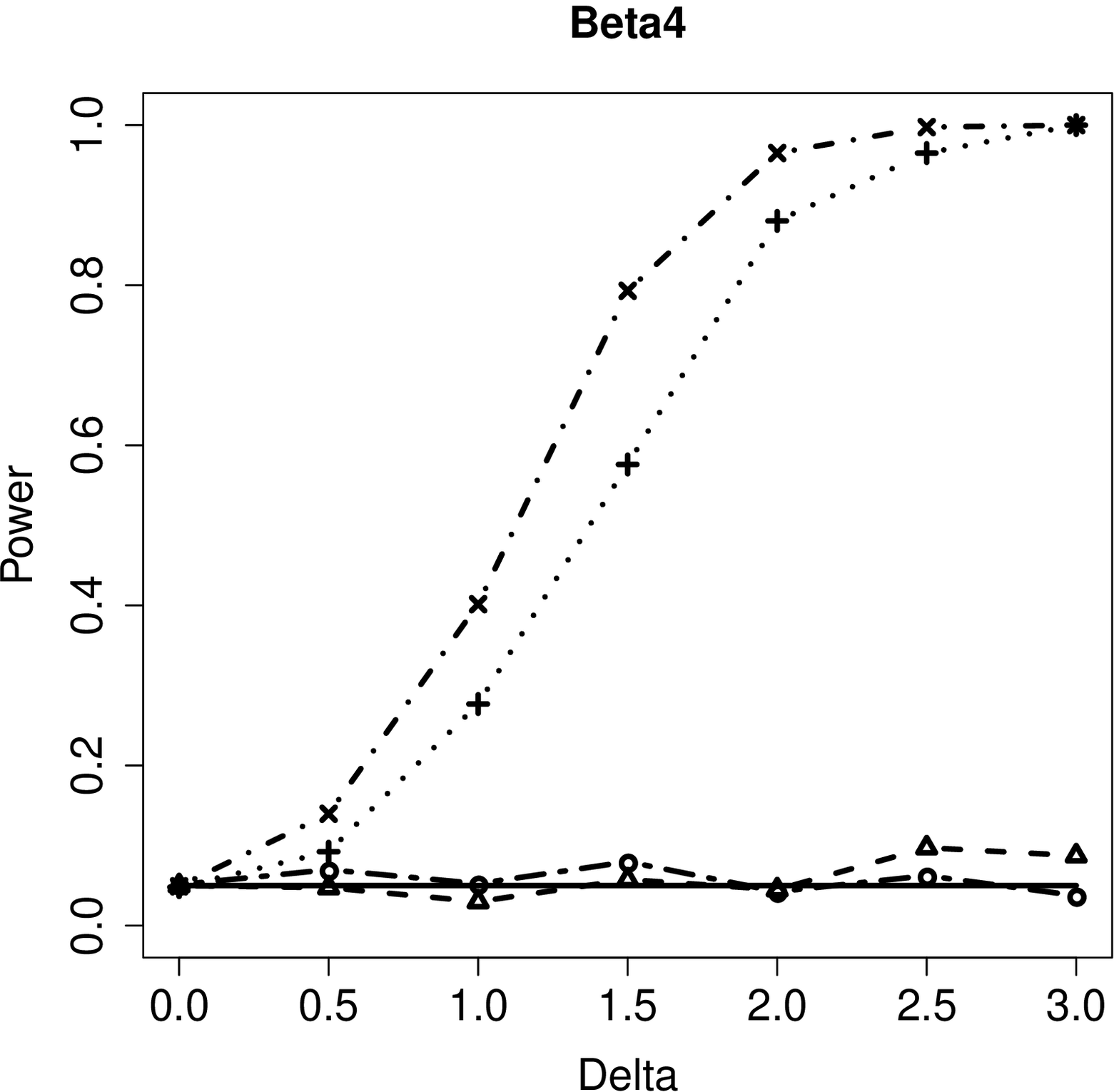}\\
&
\end{tabular}
\caption{Results illustrating the power curves from the simulation study described in section \ref{sec:sim} for $T^*$, and the F-statistic for sparsely sampled data for $\beta_1(\cdot, \cdot)$, $\beta_2(\cdot, \cdot)$, $\beta_3(\cdot, \cdot)$, and $\beta_4(\cdot, \cdot)$ (Linear kernel : dashed line with triangles, Quadratic kernel : dotted line with pluses, Gaussian kernel : dash and dot line with crosses, F-statistic : small and big dash with circles, solid horizontal line at nominal level of $\alpha, 0.05$).}
\end{figure}
%%%%%%%%%%%%%%%%%%%%%%%%%%%%%%%%
%%%%%%%%%%%%%%%%%%%%%%%%%%%%%%%%
The simulated results confirm our suppositions and suggest that the methodology developed is suitable for testing our hypothesis. 
\section{Data Analysis}
\label{sec:dat}
\cite{Kippler} found that cadmium concentration in the placenta was inversely associated with birth weight. \cite{Menai} found that maternal exposure to cadmium was inversely associated with birth weight in females. Another study detected that women with high mercury exposure delivered offsprings with reduced birth weights (\cite{Vejrup}). \cite{Xie} documented that increased maternal blood lead level was negatively related to birth weight. Other studies proved that arsenic easily crossed the placenta causing reduced birth weight(\cite{Jin}, \cite{He}, \cite{Rahman}). Zinc deficiency has been associated with low birth weights(\cite{Uriu}). Low birth weight is frequently followed by steeper growth trajectories in early life, called catch-up growth (\cite{F}, \cite{Lui}, \cite{Moore}). Such accelerated growth in early life is a consistent risk factor for cardiometabolic impairment in adulthood (\cite{Anderson}, \cite{De}, \cite{Whincup}). The hypothesis that prenatal exposures to metal increases disease risk in adulthood was first proposed by \cite{Barker} and this hypothesis has been subsequently supported by multiple epidemiologic studies (\cite{Erik}). 
We applied our proposed hypothesis testing procedure to data obtained from the Newborn Epigenetic STudy (NEST), a birth cohort study in Durham, North Carolina. 
Between $2009$ and $2011$, the NEST recruited pregnant adult women from six prenatal clinics who intended to deliver at either of the two obstetric facilities serving Durham, North Carolina, enabling collection of umbilical cord blood at birth (\cite{Hoyo}). Enrolment occurred during the first prenatal clinic visit ($\sim 13$ weeks) with questionnaire and peripheral blood collected. 
Demographic and lifestyle information such as age, race, education history, socioeconomic status, smoking during pregnancy, type of nutrients they consumed while expecting, and whether the mother breastfed her baby were collected from the pregnant ladies. At birth, cord blood was collected and processed from $619$ children and the metal levels in the blood for $24$ metals was recorded. Weight of newborn babies was collected over several years. None of the children from the study died. We looked at the data over the first five years at a monthly resolution. This led to between $2$ and $30$, randomly and irregularly spaced observations for each individual. Figure $3$ illustrates the growth curves of the children in our study.
\begin{figure}[htbp]
  \includegraphics[width=\linewidth]{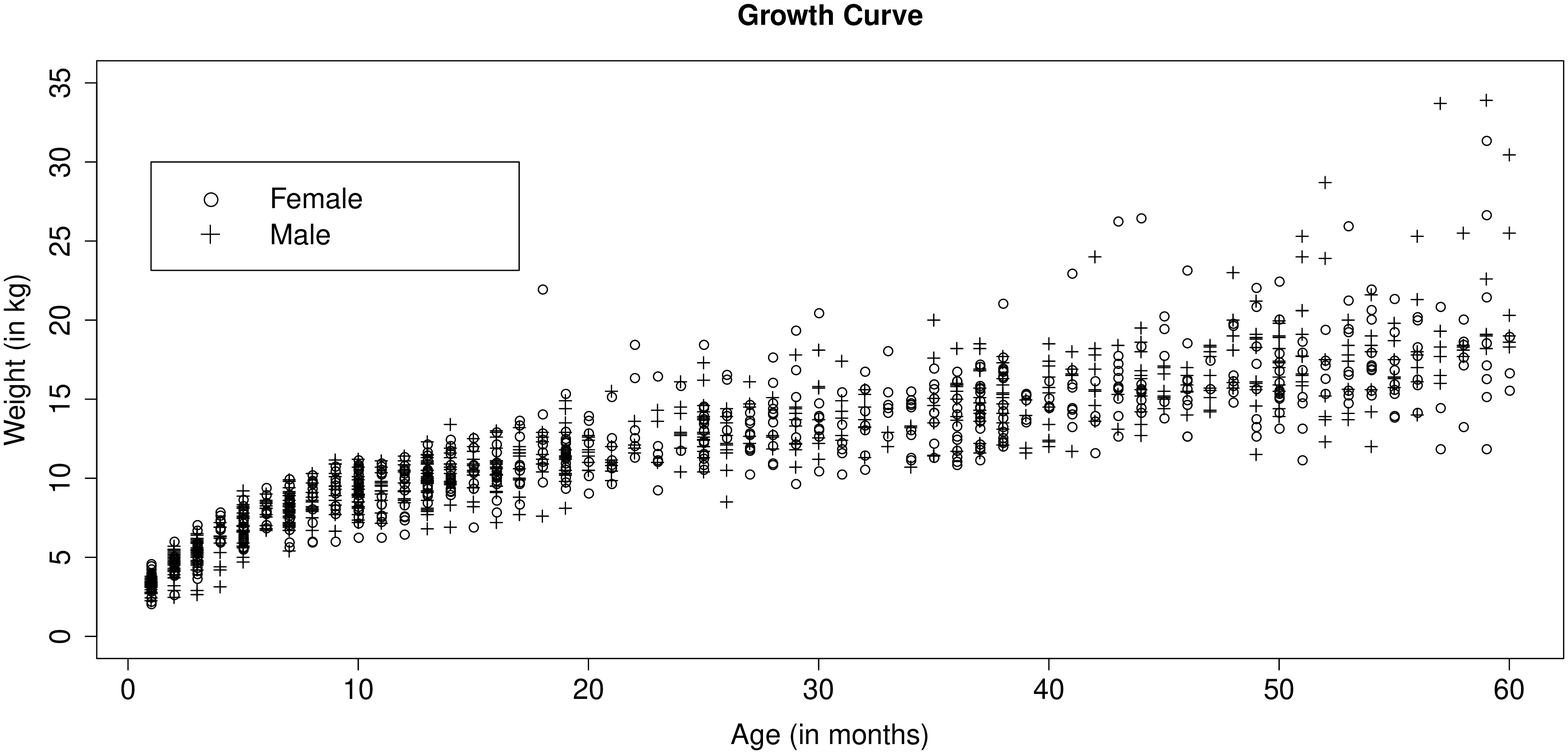}
  \caption{Growth curve showing weight of the children in kg versus their age in months for the children considered for testing for association of growth curve with metal content as described in section \ref{sec:dat}, the females being denoted by a circle, and the males being denoted by a plus.}
\end{figure}
The objective of the study was to test whether there was association between the levels of the metals in the children's cord blood and their weight curves. The weights of the children, which are functions of time, were considered as the response variables. The demographic information of the mothers were treated as the nuisance covariates. Since the underlying $\beta(\cdot, \cdot)$ function is not known, we have used the gaussian kernel in our methodology. The weight curves did not display any clear distinction between the male and female children, so the analysis was not performed separately for the genders. The first principal component function explained a large percentage of the underlying variance-covariance structure for either gender considered in testing the association of growth curve with metal content. The first eigen vector for male children, and for female children was similar to that when all the children were considered together. Given the small size of data available to us, we did not split the dataset according to gender as doing so would lead to a severe loss in power. We have included the relevant table and figures in the supplementary materials. The assumptions made in our model would be reasonable in the hypothesis test performed.
\begin{table}
\centering
%\resizebox{0.99\textwidth}{!}{\begin{minipage}{\textwidth}
\begin{tabular}[htbp]{|c|c|c|c|}
        \hline
        &\textbf{Linear} & \textbf{Quadratic} & \textbf{Gaussian}\\
        \hline
        \textbf{Number of metals} & 24 & 5 & 19 \\
        \hline
        \textbf{p-value} & 0.8937 & 0.9697 & \bf{0.0081} \\
        \hline
        \end{tabular}
        \bigskip
        \caption{Results displaying p-values for association between the metal content from the cord blood and the growth curve collected from the children in NEST mentioned in section \ref{sec:dat} for the linear, quadratic, and gaussian kernels.}
%\end{minipage}}
\end{table} 
Data about metal levels and demographic covariates both were available for $77$ children. Of these, $2$ growth curves were found to have sharp spikes, which may have been due to improper measurements, and have been left out of the analyses giving us $75$ usable records. The variables of interest were the metal levels in the cord blood. The analysis was done for all the metals together for each of the three kernels that we studied in the simulation study with $10,000$ permutations for our methodology. We are interested in simultaneously testing $3$ null hypotheses on the same dataset, so the p-values from each of the individual tests may result in false discoveries. Let the individual p-values be given by $p_1$, $p_2$, and $p_3$, respectively. The simplest approach to correct for the multiple testing is using the Bonferroni correction. We have used the Bonferroni correction and the test has been performed at a level of significance of $0.05/3 = 0.0167$. Table $2$ lists the p-values for the $3$ tests performed. The p-value of $0.0081$ for the gaussian kernel suggests that there seems to be some non-linear association between the level of metals in cord blood and the growth trajectory of children. 
\section{Discussion}
\label{sec:disc}
We have developed a new hypothesis testing method that can determine if there is a potentially non-linear effect of a multidimensional covariate on irregularly-spaced functional data. We have proposed a score-based test statistic, and applied it to the NEST dataset to test for association between the metal content in a child's cord blood and their weight curve for the first $5$ years of their life. The amount of metals in a child's cord blood seems to affect the child's weight in its early childhood. Further analyses may elucidate how the different metals interact to affect the child's growth. We ran our methodology on male and female children separately for association between metals and growth curves. 
We performed these latter tests solely for the purpose of getting a vague idea of what the p-values would look like, and shall not be using it to draw conclusions as we reused the dataset that we had used for our initial analyses.
 It may be noted that given the nature of the study, it is subjected to self-selection bias. Also, the participants were restricted to a certain geographic area. Further work may be performed for cases where certain assumptions that we made do not hold true. Our model and methodology assume that the error function is a Gaussian process. We do not know how our test statistics would behave if the assumption were to be false. We have performed a global testing for $\beta(\cdot, \cdot)$. When a $\beta(\cdot, \cdot)$ is found to be significantly different from $0$, we do not know what the underlying structure of $\beta(\cdot, \cdot)$ might be. We require localized testing in order to determine that.
  We have developed a software package in R for our methodology. When performing the hypothesis test, the true underlying function of the covariate is unknown, so it is suggested that a Gaussian kernel be used in the methodology developed. This is because if the true function is linear, using a Gaussian kernel would not cause much loss of power compared to using a linear kernel, but if the true function is a complicated non-linear one, the power obtained would be much higher compared to using a linear kernel.
 \section{Supplementary Material}
\label{sec:supp}
The code used for the simulations and the supplementary material mentioned in the data analysis section is available online at website \url{https://github.com/MBiswas1/hyp_test}. 
\bibliographystyle{biom} \bibliography{biombib}
\label{lastpage}
\end{document}

{
\begin{center}
    {\LARGE\bf Supplementary Materials for\\ Hypothesis Testing in\\ Nonlinear Function on Scalar Regression with Application to Child Growth Study}
    
    Mityl Biswas, Arnab Maity
    
      Department of Statistics, North Carolina State University
\end{center}

\newpage
\section{$\mathbf{K}$ as a Kronecker product.}
\label{sec:K}
We have utilized the following property in order to make our code more computationally efficient while performing the simulation study.
\bigskip
\\
Let's consider the $(h, k)^{th}$ element of the $(i,j)^{th}$ block for some $h \epsilon \{1,$ $\ldots,$ $m_i\}$, $k \epsilon \{1,$ $\ldots,$ $m_j\}$, $i \epsilon \{1,$ $\ldots,$ $n\},$ $j \epsilon \{1,$ $\ldots,$ $n\}.$
\begin{center}
$K\{(\mathbf{X}_i, \mathbf{t_{ih}}), ( \mathbf{X}_j, \mathbf{t_{jk}})\} = L(\mathbf{X_i}, \mathbf{X_j})e^{-|t_{ih}-t_{jk}|},$
\end{center}
$h \epsilon \{1,$ $\ldots,$ $m_i\}$, $k \epsilon \{1,$ $\ldots,$ $m_j\}$, $i = 1, \ldots, n,$ $ j = 1, \ldots, n$. 
This is clearly a separable function in $(\mathbf{X}_i, \mathbf{X}_j)$, and $(t_{ih}, t_{jk})$. Then, we can see that the $(i,j)^{th}$ block is $L(\mathbf{X}_i, \mathbf{X}_j) \otimes \mathbf{T_{ij}}$, where $\mathbf{T_{ij}}$ is the $m_i \times m_j$ matrix with the $(h, k)^{th}$ element as $e^{-|t_{ih}-t_{jk}|}$ for $h = 1, \ldots, m_i, k = 1, \ldots, m_j$. 
Now, if we had $m_i = m$, with $t_{ih} = t_h$ for $h = 1, \ldots, m, i = 1, \ldots, n$, then, the $(i,j)^{th}$ block would have been $L(\mathbf{X}_i, \mathbf{X}_j) \otimes \mathbf{T}$, where $\mathbf{T}$ is the $m \times m$ matrix with the $(h, k)^{th}$ element as $e^{-|t_{h}-t_{k}|}$ for $h = 1, \ldots, m, k = 1, \ldots, m$. This results in the following corresponding kernel matrix: $\mathbf{K^*} = \mathbf{A} \otimes \mathbf{T}$, where $\mathbf{A}$ is the $n \times n$ matrix with the $(i, j)^{th}$ element as $L(\mathbf{X_i}, \mathbf{X_j})$. Now, if we only consider those rows and columns of $K^*$ that correspond to the observed data, we would have the required kernel matrix $\mathbf{K}$.

\section{Additional Tables and Figures}
\label{sec:sup}
The first principal component function explained a large percentage of the underlying variance-covariance structure for either gender considered in testing the association of growth curve with metal content. This is evident in table 1. The first eigen vector for male children, and for female children was similar to that when all the children were considered together. This is illustrated in figure 1.\\
Table 1 here.\\
%Table 2 here.\\
Figure 1 here.\\
%Figure 2 here.
The p-values obtained by performing the tests separately for the male and the female children show that they are similar for the two genders, although the values are slightly lower for the females. The statistics are listed in table 2.\\
Table 2 here.\\
The p-values obtained by performing the tests separately for each of the $24$ metals have been recorded in table 3. We obtained these numbers solely for the purpose of getting a vague idea of what the p-values would look like, and shall not be using them to draw conclusions as we have reused the dataset that we had used for our initial analyses.\\
Table 3 here.
}

\begin{table}
\centering
%\resizebox{0.99\textwidth}{!}{\begin{minipage}{\textwidth}
\begin{tabular}[htbp]{|c|c|c|c|}
        %\hline
       % \multicolumn{4}{|c|}{\textbf{Study regarding methylation at DMR sites}} \\
        %\hline
        %&\textbf{All} & \textbf{Female} & \textbf{Male}\\
        %\hline
         %\textbf{Number of children} & 59 & 30 & 29\\
        %\hline
        %\textbf{Explained by first principal component function} & 95.0 \% & 98.1 \% & 90.6 \% \\
        %\hline
        %\textbf{Explained by second principal component function} & 3.6 \% & 1.9 \% & 5.9 \% \\
        %\hline
        %\multicolumn{4}{|c|}{\textbf{Study regarding metal content in cord blood}} \\
        \hline
        &\textbf{All} & \textbf{Female} & \textbf{Male}\\
        \hline
         \textbf{Number of children} & 75 & 38 & 37\\
        \hline
        \textbf{Explained by first principal component function} & 95.4 \% & 98.1 \% & 99.9 \% \\
        \hline
        \textbf{Explained by second principal component function} & 3.2 \% & 1.9 \% & 0.1 \% \\
        \hline
        \end{tabular}
        \caption{Percentage of the underlying variance-covariance structure of $\Sigma$ explained by the first $2$ principal component functions for the individuals with and without segregation by their gender for studying the association of growth curve with the metal content in cord blood as mentioned in the data analysis section.}
%\end{minipage}}
\end{table} 
%\newpage
%\begin{figure}[htbp]
%\centering
%  \includegraphics[width=\linewidth]{evec_gen.pdf}
% \caption{Eigen vectors corresponding to the first principal component for studying the association of growth curve with the methylation at DMR sites for all subjects (solid line), females (dashed line), and males (dotted line) as mentioned in the data analysis section.}
%\end{figure}

\newpage
\begin{figure}[htbp]
  \includegraphics[width=\linewidth]{evec_met2.pdf}
  \caption{Eigen vectors corresponding to the first principal component for studying the association of growth curve with the metal content in cord blood for all subjects (solid line), females (dashed line), and males (dotted line) as mentioned in the data analysis section.}
\end{figure}

\newpage

%\resizebox{0.99\textwidth}{!}{\begin{minipage}{\textwidth}
\begin{table}
\centering
\resizebox{\textwidth}{!}
{\begin{minipage}{\textwidth}
\centering
\begin{tabular}[htbp]{|c|c|c|c|}
        \hline
        &{\textbf{All Metals}}  & {\textbf{Toxic Metals}} & {\textbf{Other Metals}}\\
        \hline
        \textbf{Females} & 0.0163 & 0.2019	& 0.0174\\
        \hline
        \textbf{Males} & 0.0282 & 0.2882	& 0.0303\\
        \hline
        \end{tabular}
        \caption{The p-values (rounded to fourth decimal place) for the test of association between growth curves and amount of metal in a child's cord blood for either gender as mentioned in the discussion section.}
\end{minipage}}
\end{table} 

\newpage

\begin{table}
\centering
%\resizebox{0.99\textwidth}{!}{\begin{minipage}{\textwidth}
\begin{tabular}[htbp]{|c|c|c|c|}
        %\hline
       % \multicolumn{4}{|c|}{\textbf{Study regarding methylation at DMR sites}} \\
        %\hline
        %&\textbf{All} & \textbf{Female} & \textbf{Male}\\
        %\hline
         %\textbf{Number of children} & 59 & 30 & 29\\
        %\hline
        %\textbf{Explained by first principal component function} & 95.0 \% & 98.1 \% & 90.6 \% \\
        %\hline
        %\textbf{Explained by second principal component function} & 3.6 \% & 1.9 \% & 5.9 \% \\
        %\hline
        %\multicolumn{4}{|c|}{\textbf{Study regarding metal content in cord blood}} \\
        \hline
        &\textbf{All} & \textbf{Female} & \textbf{Male}\\
        \hline
         \textbf{Number of children} & 75 & 38 & 37\\
        \hline
        \textbf{Cadmium} & 0.4472 & 0.3489 & 0.4136 \\
        \hline
        \textbf{Lead} & 0.7191 & 0.4567 & 0.1715 \\
        \hline
        \textbf{Arsenic} & 0.7102 & 0.6798 & 0.3210 \\
        \hline
        \textbf{Iron} & 0.4734 & 0.0998 & 0.8637 \\
        \hline\textbf{Zinc} & 0.2612 & 0.6560 & 0.7337 \\
        \hline
        \textbf{Selenium} & 0.8298 & 0.0694 & 0.7281 \\
        \hline\textbf{Copper} & 0.9028 & 0.7353 & 0.0164 \\
        \hline
        \textbf{Calcium} & 0.2104 & 0.1559 & 0.3459 \\
        \hline\textbf{Strontium} & 0.6351 & 0.0676 & 0.7209 \\
        \hline
        \textbf{Barium} & 0.6884 & 0.9244 & 0.9779 \\
        \hline\textbf{Manganese} & 0.4178 & 0.8938 & 0.1069 \\
        \hline
        \textbf{Cobalt} & 0.5255 & 0.0657 & 0.9518 \\
        \hline\textbf{Molybdenum} & 0.9491 & 0.3438 & 0.9059 \\
        \hline
        \textbf{Scandium} & 0.6514 & 0.3574 & 0.6255 \\
        \hline\textbf{Titanium} & 0.4075 & 0.2335 & 0.5048 \\
        \hline
        \textbf{Chromium} & 0.2799 & 0.5231 & 0.5076 \\
        \hline\textbf{Aluminium} & 0.9435 & 0.8885 & 0.3141 \\
        \hline
        \textbf{Silicon} & 0.3530 & 0.0963 & 0.6416 \\
        \hline\textbf{Vanadium} & 0.4860 & 0.1063 & 0.1957 \\
        \hline
        \textbf{Rubidium} & 0.4089 & 0.8445 & 0.2812 \\
        \hline\textbf{Tin} & 0.4601 & 0.0135 & 0.2599 \\
        \hline
        \textbf{Mercury} & 0.4742 & 0.4186 & 0.4358 \\
        \hline\textbf{Nickel} & 0.5029 & 0.1059 & 0.3895 \\
        \hline
        \textbf{Magnesium} & 0.7958 & 0.2419 & 0.3440 \\
        \hline
        \end{tabular}
        \caption{The p-values (rounded to fourth decimal place) for the test of association between growth curves and amount of metal in a child's cord blood for both genders combined, and either gender separately for each of the $24$ metals whose values are available to us, as mentioned in the discussion section.}
%\end{minipage}}
\end{table}